\documentclass[letterpaper,twocolumn,10pt]{article}

\usepackage{usenix2019_v3}
\usepackage[tight,footnotesize]{subfigure}
\usepackage{url}
\usepackage{soul}
\usepackage{graphicx}
\usepackage[utf8]{inputenc}
\usepackage{xspace}
\usepackage{booktabs}
\usepackage{bookmark}
\usepackage{amsmath}
\usepackage{cleveref}
\usepackage{siunitx}
\usepackage{flushend}
\usepackage{hyperref}
\usepackage{textcase}
\usepackage{textcomp}
\usepackage[english]{babel}
\usepackage[toc,page]{appendix}
\usepackage[ruled,vlined]{algorithm2e}

\newcommand{\eg}{e.g.,\xspace}

\newcommand{\etal}{\textit{et~al.\xspace}}

\newcommand{\ie}{i.e.,\xspace}

\newcommand{\unit}[1]{\ensuremath{\, \mathrm{#1}}}
\newcommand{\ignore}[1]{}
\newcommand{\recvname}{\NoCaseChange{SemperFi\xspace}}
\newcommand{\pseurec}{pseudorange rectifier}
\newcommand{\pseurecuc}{Pseudorange Rectifier}

\newcommand{\api}{adversarial peak identifier\xspace}
\newcommand{\apiuc}{Adversarial Peak Identifier\xspace}

\newcommand{\rss}{legitimate signal retriever}
\newcommand{\rssuc}{Legitimate Signal Retriever}
\graphicspath{{./figures/}}

\begin{document}
\title{\recvname{}: A Spoofer Eliminating GPS Receiver for UAVs}

\author{
{\rm Harshad Sathaye, Gerald LaMountain, Pau Closas, and Aanjhan Ranganathan}\\
Northeastern University, Boston, MA, USA
}

\maketitle

\begin{abstract}
It is well-known that GPS is vulnerable to signal spoofing attacks.
Although several spoofing detection techniques exist, they are incapable of mitigation and recovery from stealthy attackers.
In this work, we present \recvname{}, a single antenna GPS receiver capable of tracking legitimate GPS satellite signals and estimating the true location even during a spoofing attack. 
The main challenge in building \recvname is, unlike most wireless systems where \emph{the data} contained in the wireless signals is important, GPS relies on the time of arrival (ToA) of satellite signals. 
\recvname{} is capable of distinguishing spoofing signals and recovering legitimate GPS signals that are even completely overshadowed by a strong adversary.
We exploit the short-term stability of inertial sensors to identify the spoofing signal and extend the successive interference cancellation algorithm to preserve the legitimate signal's ToA.
We implement~\recvname in GNSS-SDR, an open-source software-defined GNSS receiver and evaluate its performance using UAV simulators, real drones, a variety of real-world GPS datasets, and various embedded platforms.
Our evaluation results indicate that in many scenarios, \recvname can identify adversarial peaks by executing flight patterns that are less than 50 m long and recover the true location within 10 seconds (Jetson Xavier).
We show that our receiver is secure against stealthy attackers who exploit inertial sensor errors and execute seamless takeover attacks.
We design \recvname{} as a pluggable module capable of generating a spoofer free GPS signal for processing on any commercial-off-the-shelf GPS receiver available today.
Finally, we release our implementation to the community for usage and further research.

\end{abstract}

%\pagenumbering{gobble}

\section{Introduction}
A wide variety of applications such as positioning, navigation, asset and personnel tracking, communication systems, power grids, emergency rescue and support, and access control use Global Positioning System (GPS) ubiquitously to estimate location and time. Given the popularity of unmanned vehicular systems such as self-driving cars, GPS application in safety- and security-critical applications is increasing. 

Due to the lack of authentication in civilian navigation messages, GPS is vulnerable to signal spoofing attacks. 
In a GPS signal spoofing attack, the attacker transmits specially crafted signals that imitate satellite signals with power high enough to overshadow the legitimate signals~\cite{amin2016vulnerabilities}. 
Several researchers have shown that it is possible to modify the course of ships~\cite{texas2013yachtspoofing}, unmanned aerial vehicles~\cite{shepard2012drone} and self-driving cars~\cite{tesla2019hack} by simply spoofing GPS signals. 
There is also an increase in GPS spoofing incidents~\cite{c4ads2019spoofingreport} reported from around the world. 
For example, there are reports of thousands of ships in Shanghai falling victim to GPS spoofing~\cite{shanghai2019gpshack}. 
There are also reports~\cite{c4ads2019spoofingreport} of state actors using GPS spoofing and jamming in several countries to disrupt everyday affairs. 
With the widespread availability of software-defined radio and public GPS signal generator repositories~\cite{osqzss2015gpssim}, it is now possible to spoof GPS signals with less than \$100 of hardware equipment.
Furthermore, it is possible to use GPS spoofing to trip power generators in smart grids triggering false activation of automatic control systems and potentially leading to wide-area power blackout~\cite{risbud2018vulnerability}. 

Proposed countermeasures are either cryptographic solutions or leverage physical-layer signal properties. 
Countermeasures that use some form of cryptographic authentication~\cite{kuhn2004asymmetric, wesson2012practical, lo2010authenticating, cheng2009authenticity} prevent attackers from generating arbitrary false GPS signals. 
However, they do not protect against attackers capable of recording and replaying legitimate GPS signals. The receiver's location and time are estimated using the GPS signal's time-of-arrival and not just on the navigation message content. 
Other countermeasures that do not require cryptographic authentication rely on detecting anomalies in the received GPS signal's physical characteristics, such as received signal strength~\cite{warner2003gps}, noise levels, direction or angle of arrival~\cite{meurer2016direction}, and other data that are readily available as receiver observables on many COTS GPS receivers. 
Some countermeasures~\cite{ranganathan2016spree} exploit the difficultly in canceling out legitimate GPS signals completely to detect stealthy, seamless takeover attackers. 
A few countermeasures propose the use of additional sensors~\cite{jafarnia2012detection} and receivers~\cite{tippenhauer2011requirements,montgomery2011receiver} to detect spoofing attacks. 
The majority of the above schemes only detect a GPS spoofing attack, i.e., raise the alarm in case of a spoofing attack and often require manual intervention, unlike SemperFi.
Moreover, existing spoofing mitigation techniques are ineffective against strong adversaries capable of completely overshadowing legitimate signals and stealthy attackers, e.g., seamless takeover~\cite{tippenhauer2011requirements} of victim's GPS location without any signal disruption despite having redundant fail-safe sensors~\cite{narain2019security}. 
In summary, today's GPS receivers, specifically those implemented on UAVs, are incapable of uninterrupted operation during a spoofing attack.

In this work, we present \recvname, a single-antenna GPS receiver for UAVs that autonomously recovers and continues to output legitimate location during a spoofing attack. 
\recvname comprises of three main building blocks: i) Spoofing detector, ii) Adversarial Peak Identifier (API), and iii) \rssuc{} (LSR). 
The spoofing detector provides reliable spoofing detection.
The API is responsible for detecting a spoofing attack \emph{and}, distinguishing the attacker's signal from the legitimate GPS signals.
If necessary, the LSR synthesizes an appropriate recovery signal, and eliminates the spoofing signal using a successive interference cancellation (SIC) technique. 
Once spoofing is detected, the \api{} module instructs the drone to perform a randomly generated maneuver that lasts for about 10-20 s. 
\recvname exploits the short-term stability of IMU sensors and correlates its pattern estimates with that of the GPS.
Adversarial and legitimate signals are identified based on correlation results. Peak identification information is then passed on to the \rss{}.
Traditional wireless communication systems have successfully applied SIC to recover message contents. It is important to note that, in the case of GPS, in addition to the data contained within the navigation messages, it is essential to preserve the ToA of the satellite signal itself. 
To address this unique challenge present in eliminating GPS spoofing signals, we develop algorithms to estimate the various physical characteristics such as amplitude, phase, and ToA of both the legitimate and spoofing signals. 

We implement \recvname using GNSS-SDR~\cite{fernandez2011gnss} and evaluate its performance against both synthetically generated as well as real-world GPS signals using consumer drones like DJI Flamewheel F450~\cite{djiflamewheel} and Holybro S500~\cite{holybros500}. %-standard UAV simulators such as Gazebo~\cite{gazebo} and ArduPilot~\cite{ardupilot}.
We also evaluate the performance of \recvname on various embedded systems commonly used as UAV flight controllers.
Furthermore, we evaluate the effectiveness of \recvname against TEXBAT~\cite{humphreys2012texas}, a public dataset of GPS spoofing traces. 
Our evaluation show that in the majority of attack scenarios, SemperFi can recover with an accuracy of less than 20 m.
Moreover, performance evaluation on popular platforms like Jetson Nano and Xavier take less than 10 s to recover the legitimate location.
It is also possible to deploy \recvname as a pluggable module that outputs spoofer-free GPS signals identical to legitimate satellite signals. 
Therefore, \recvname allows an unmodified COTS GPS receiver to process and generate location and time estimates without any disruption.
\section{GPS and Spoofing Attacks}
\label{sec:background}

\subsection{GPS Overview}
Global Positioning System (GPS) is the most widely used Global Navigation Satellite System (GNSS) that uses the L1 frequency band.
GPS consists of 31 operational satellites at an altitude of approximately 20,220 km\footnote{As of January 2021. 
\cite{gps2020gov}}.
Each satellite continuously transmits navigation messages containing timing information, satellites' ephemeris data, and other necessary information that enables the receiver on the ground to localize itself. 
The navigation messages are spread using a coarse-acquisition (C/A) code unique for each satellite and transmitted using a $1575.42\unit{MHz}$ carrier. 
The C/A code is public and contains $1023$ bits (also referred to as \emph{chips}) repeated every $1\unit{ms}$. 
Military GPS signals use a longer and a secret spreading code.
This paper focuses on civilian GPS signals as they are widely used even in security-critical applications~\cite{silverstein2016electric, goldstein2013gps}.
The navigation data comprises of five subframes. Each subframe contains $1500\unit{bits}$ at $50\unit{bps}$~\cite{borre2007software}.
These subframes contain satellite clock informatio and satellite orbital information. 
The ephemeris data is updated every 2 hrs and is valid for 4 hrs~\cite{dunn2012global}. \\

\noindent A typical GPS receiver consists of four main components, i) RF front end, ii) Acquisition module, iii) Tracking module, and iv) Position Velocity Time (PVT) module. 

\noindent\textbf{RF front-end} receives raw RF signals and converts the raw signal to an intermediate frequency for efficient processing. 
Each satellite is assigned a ``channel''. 
This channel is similar to a hardware pipeline for processing a single satellite.

\noindent\textbf{Acquisition module} performs a two-dimensional search for visible satellite signals in the received signal by correlating the received signal with a locally generated replica of each satellite's C/A code. 
A two-dimensional search is a time-domain and frequency-domain search to account for code phsae delay and Doppler shift that arise because of the satellite's and the receiver's motion.
If the code and Doppler searches result in a correlation peak above a certain threshold, the receiver then switches to tracking and demodulating the navigation message data. 

\noindent \textbf{Tracking module} is responsible for tracking the code phase and the Doppler shift provided by the acquisition module.
It also demodulates the navigation messags and passes on to the PVT module.

\noindent \textbf{Position Velocity Time Estimation (PVT)} module decodes raw navigation bits and calculates pseudorange between the satellite and the receiver. 
A receiver requires information from at least four satellites for accurately calculating position, velocity, and time.
The PVT module is the last block of the GPS receiver and implements algorithms to compute navigation solutions and delivers information in appropriate formats (\eg RINEX, UBX, NMEA~\cite{navmsgformat}) for further processing.

\subsection{Attacker goals and assumptions}
\label{subsec:spoofing_attacks}
In a GPS spoofing attack, an adversary transmits specially-crafted radio signals identical to authentic GPS satellite signals.
The spoofing signals are generated for an attacker-defined trajectory or static position and transmitted typically using a software-defined radio hardware platform.
All the necessary information for generating GPS signals like modulation schemes, message formats, and spreading codes is public knowledge.
The goal of an attacker can be to i) force the user to calculate a wrong geographic location, ii) forge timing information, or iii) execute a denial of service attack by causing interference.
During a spoofing attack, the GPS receiver locks 
onto the stronger signal \ie the attacker's signals, ignoring the weaker legitimate satellite signals.
This results in the receiver computing a false position, velocity, and time-based on the spoofing signals.
Note that the received GPS signal power on the ground is typically around $-127.5\unit{dBm}$ and, therefore, trivial for an attacker to overshadow the legitimate signal with the spoofing signal.

In this work, we focus on an attacker that forces the user to calculate a wrong geographic location. 
We do not consider an attacker whose goal is to cause a denial of service attack by transmitting jamming signals.
An attacker can manipulate the calculated PVT solution in two ways: i) manipulate ToA of messages or ii) manipulate navigation message contents (\eg{ satellite location, transmission time}). 
We base the attacker model on work done in~\cite{tippenhauer2011requirements} and drone hijacking strategies proposed in~\cite{noh2019tractor}.
We assume the following about the attacker. 
The attacker can have omnidirectional or directional antennas. 
They can spoof any number of satellites. 
We do not restrict the position of the attacker. 
The attacker is aware of SemperFi and can craft spoofing signals accordingly. 
We assume that the attacker has not compromised the onboard sensors and that these sensors provide valid, unadulterated data.
The attacker can execute a spoofing attack using any of the methods mentioned earlier. 
The attacker is capable of executing the sophisticated seamless-takeover attack as described in~\cite{tippenhauer2011requirements}.
In this attack, the receiver does not undergo abrupt loss of signal reception or lock. 
In a seamless takeover attack, the attacker keeps the navigation message content identical to the legitimate GPS signals and gradually introduces offsets in the code phase delays, affecting the pseudorange calculations. 
The most popular way of executing such spoofing attacks is to use GPS signal generators (both hardware~\cite{labsat} or software~\cite{osqzss2015gpssim}) to generate the spoofing signals.

Our proposed GPS receiver, \recvname, can counteract all the types of attackers mentioned above. We focus specifically on stealthy seamless takeover attacks and, in general, attacks that are hard to not only detect but pose challenges to realizing a fully-autonomous GPS receiver capable of uninterrupted true location estimates even in an adversarial setting.
\section{Design of \recvname}

\recvname is a single-antenna GPS receiver capable of providing uninterrupted location estimates even when subjected to a stealthy GPS spoofing attack.
In this section, we present the design of \recvname{} and the challenges that follow.

\subsection{Challenges}
For the GPS receiver to operate autonomously in adversarial signals, the receiver must continuously perform the following actions.
First, it is necessary to detect an ongoing spoofing attack reliably.
Then, the receiver must be capable of identifying or distinguishing between spoofing signal and the legitimate signal. 
Finally, after identifying the spoofing signal, the receiver has to eliminate or reduce the spoofing signal's effect on the final estimated location.
Unlike typical wireless communication systems where it is sufficient to recover the signals' data, GPS receivers require both the data and the ToA of the signal.
Moreover, unlike typical systems, GPS receivers are not tolerant to received sample losses. Continuous tracking of the satellite signals is necessary to estimate code and carrier phase delays that directly affect the PVT estimation.
Finally, in the case of a spoofing attack that injects fake dynamic motion pattern (e.g., diverting the course of a ship or force a drone to deviate from its flight path), the attacker dynamically manipulates the ToA of the spoofing signals as well as the data contained within the navigation messages. 
Therefore, traditional interference cancellation and mitigation techniques need to be modified or extended in order to handle this kind of attack. 

\subsection{High-level Overview} 
\begin{figure}[t]
    \centering
    \includegraphics[width=\columnwidth]{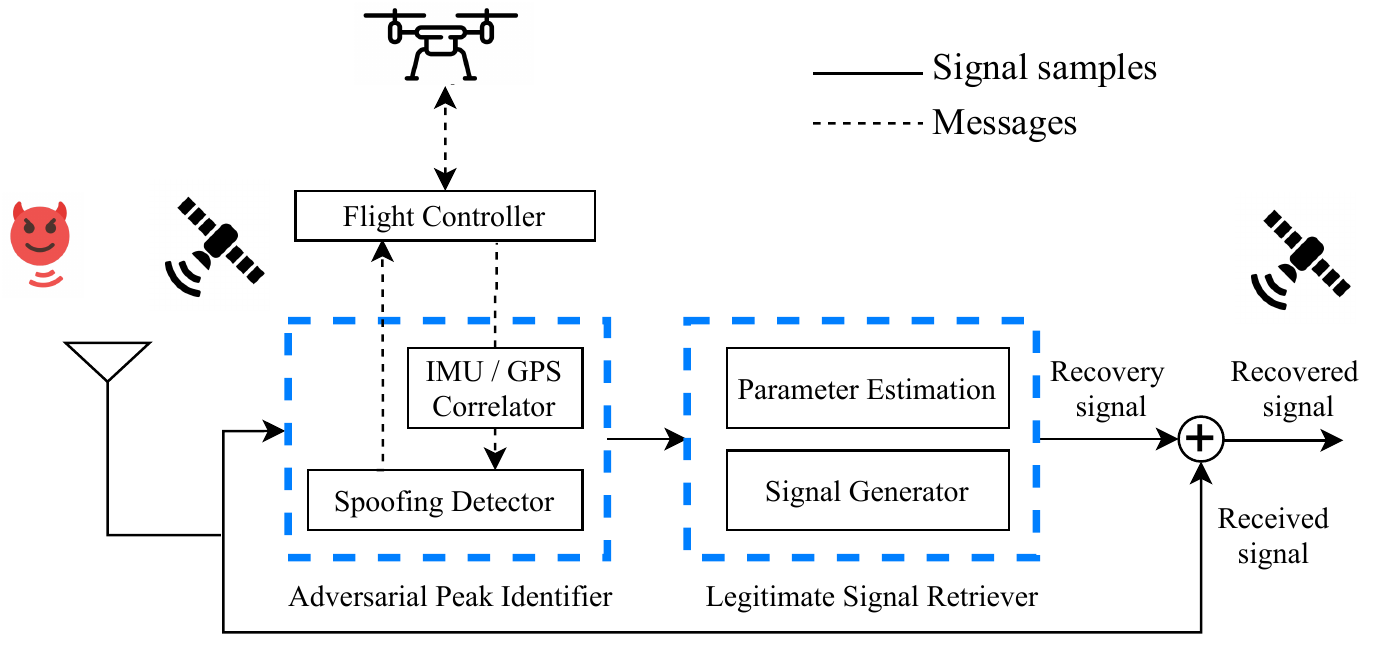}
    \caption{High-level overview depicting essential components of \recvname.}
    \label{fig:gps_receiver_ase}
\end{figure}

\recvname provides fully-autonomous spoofing resistance through the combined effort of three modules: i) Spoofing Detection, ii) the Adversarial Peak Identifier (API), and finally, iii) \rssuc{} (LSR).
A block diagram of \recvname's various components is shown in~\Cref{fig:gps_receiver_ase}.
Several spoofing detection techniques can be integrated into \recvname as long as they provide reliable spoofing signal detection for all possible adversaries. 
In this work, the spoofing detection methodology employed is based on the design of a prior work~\cite{ranganathan2016spree} that demonstrated the ability to detect even a strong, seamless takeover attack. 
Peak identification and signal recovery modules are activated in case spoofing is detected. 
The receiver continuously analyses the incoming signal for a spoofing attack and raises an alarm once spoofing is detected.
Upon spoofing detection, the API performs a pseudorandom maneuver to identify the adversarial peak.
It correlates position estimates obtained from inertial measurement unit (IMU) data and GPS to accurately identify if the currently tracked signals are adversarial.
LSR is generates a replica of the adversarial signal and performs SIC to recover the legitimate signal. \\

\noindent
SIC is a technique used for canceling out interference caused by stronger signals. 
The GPS signal from a single satellite can be modelled as
    \begin{align}
        S_{R} = a[k]\tilde{s}_{T}[k-\tau(k)]e^{j2\pi f_{D}[k]T_{s}k + \phi[k]}
    \end{align}
where $s_{T}(k)$ is the baseband signal ($k$ number of samples per C/A code), and $a[k], \tau(k), f_{D}[k],\phi[k]$ are the amplitude, time-varying code delay, Doppler shift and carrier phase shift respectively.
In presence of an adversary, received signal is
    \begin{align}
        S_{R} = S_{L} + S_{AT}
    \end{align}
where $S_{L}$ is the legitimate signal and $S_{AT}$ is the attacker's signal. In a GPS spoofing attack, the attacker overpowers the legitimate signal. Thus, $a_{AT} > a_{L}$ and as result, the GPS receiver tracks $S_{AT}$.
The LSR module uses the spoofing detector's tracking parameters $\tau_{AT},$ Doppler shift $f_{AT},$ to track the adversary signal for a specific duration and extracts the baseband data $s_{AT}$. 
The amplitude $a_{AT}$ and carrier phase shift $\phi_{AT}$ of the adversarial signal are then estimated and used in combination with the baseband data to generate the recovery signal $S^{'}_{AT}$, a close replica of the estimated adversary signal.
Using the above information, $S_{L}$ can be obtained as follows:
    \begin{align}
        S_{L} = S_{R} - S^{'}_{AT}
    \end{align}
The replica is fed back to perform SIC and undergoes re-acquisition. 
If necessary, \recvname repeats this process until the spoofing detector does not trigger an alarm. 
At this stage, the spoofing signal is eliminated or significantly attenuated, and therefore the receiver starts tracking the legitimate signals. 
There are scenarios where, despite a successful recovery, either due to the spoofing signal's strength or synchronization concerning the legitimate signals, the navigation message content and arrival time are hard to decode and introduce ambiguities in the PVT estimates.  
We developed a pseudorange rectifier for such specific scenarios that can recover from an attack with decreased accuracy.
Finally, we designed \recvname as a plugin module that can be configured to act as a spoofing signal filter, where the filtered signal is fed directly to any commercial GPS receiver for PVT estimation. 
This prevents significant hardware design changes to existing deployments.

\label{sec:high_overview}
\subsection{Adversarial Peak Identifier (API)}
The presence of a valid satellite signal is determined by a peak that forms as a result of the correlation operation performed by the acquisition module.  
Malicious signals result in additional correlation peaks which may be misidentified as legitimate GPS signals.
The adversarial peak identifier (API) is responsible for identifying such malicious peaks.
By nature, just like every wireless receiver, the GPS receiver also locks on to the strongest signal and tracks it.
Thus, even in scenarios where the receiver receives both adversarial and legitimate signals, it calculates the stronger GPS signals' PVT solution.
The spoofing detection strategy that we deploy will declare spoofing based on multiple peaks in the acquisition plot.
\recvname then attempts to attenuate adversarial signals to recover from the spoofing attack.
This is not, however, a simple matter of attenuating the signal producing the strongest peak.

An attacker aware of this strategy can transmit signals with a power lower than the received signal in specific attack scenarios.
Even though the attacker's signal is weaker, it will still be visible in the acquisition plot.
As a result, the spoofing detector will declare positive spoofing.
If the stronger peak is assumed to be the adversarial peak, then \recvname{} will eliminate the legitimate peak as the legitimate signal is stronger than the adversarial signal.
Therefore, for \recvname to successfully attenuate adversarial signals and recover the location, it is essential to identify the adversarial peak from the acquisition plot in a way that accounts for these scenarios.

In our design, the API identifies adversarial peaks based on the following procedure.
Once spoofing is detected, API signals the UAV to stop and hold the current position. 
As the UAV stops, the spoofed signals' location should also reflect this stop in an ideal attack scenario. 
This is possible as the attacker knows how the UAV is supposed to move based on spoofed trajectories. 
Once the UAV stops, the UAV performs a specific maneuver consisting of a pseudorandom sequence of turns. 
The attacker is not aware of the exact maneuver and cannot generate GPS signals that reflect this maneuver.
Before performing the maneuver, UAV de-couples GPS and extended Kalman filter (EKF) and uses IMU sensors based dead-reckoning to track the movements.
The UAV logs GPS coordinates, but it does not use it for rectifying EKF estimates.
API compares the track estimated by IMU sensors and GPS by averaging the Euclidean distance between each deviation sample obtained from IMU sensors and the GPS receiver. 
Since the attacker cannot generate the signals corresponding to the maneuver, a comparison of track estimates by IMU sensors and GPS shows significant deviations. 
This confirms that the GPS receiver is locked on to adversarial peaks. 
API relays this information to LSR, which then attenuates the peak.
On the contrary, if the comparison does not show deviations, it is concluded that the receiver is tracking the legitimate peak even if there is an ongoing spoofing attack.

\label{sec:api}
\subsection{Legitimate Signal Retriever (LSR)}
LSR is responsible for generating the corresponding replica signal \ie the recovery signal for every spoofed satellite.
LSR requires; i) Amplitude, ii) code phase delay, iii) Doppler shift, iv) carrier phase, and v) navigation bit of the attacker's signal for generating the recovery signal.
LSR obtains the code phase delay and the Doppler shift from the acquisition module.
The replica signal is aligned with the received spoofing signal in the time domain using the code phase delay and the frequency domain using the Doppler shift.
The LSR consists of a minimal tracking module that extracts the navigation bits and the carrier phase information of the adversarial spoofing signal. 
Each of the required components except the signal amplitude is readily available through the basic acquisition and tracking components in any standard receiver architecture.
We devised an amplitude estimation technique that relies on the correlation coefficient of the attacker's peak.\bigskip

\noindent \emph{Amplitude Estimation:} The amplitude of the acquired signal can be estimated from the magnitude of the corresponding peak in the two-dimenstional function of code phase delay and the Doppler shift called the cross-ambiguity function (CAF).
Recall that the input to the acquisition block is a set of $K$ observations of a modulated GNSS signal. The sampled baseband signal can be modeled as
    \begin{align}
        x_{IN}[k] = a[t]\tilde{s}_{T}[t-\tau(t)]e^{j2\pi f_{D}[t]T_{s}t + \phi[t]}
    \end{align}
    where $a[t]$ is the signal amplitude, $\tilde{s}_{T}[t]$ is a filtered and sampled version of the complex baseband GNSS signal. Computation of the correlations which comprise the sampled CAF, in the acquisition block is typically done in the Fourier domain after carrier wipe-off.
    \begin{align}
        x[k] = x_{IN}[k]\cdot e^{-j2\pi \check{f}_{D} tT_{s}}
    \end{align}
    At the peak of the CAF, the parameters $\check{f}_{D}[t],\check\tau[t],\check\phi[t]$ correspond to the maximum likelihood estimate of the ``true'' parameter values, and the discrete Fourier domain representation of the signal after wipe-off simplifies to
    \begin{align}
        X[k] = \textsc{FFT}_{K}{\{x[k]\}} = a[t]*S[t]W_{K}^{\tau}
    \end{align}
    Applying the FFT of the local code replica $D[k]$ is performed by multiplication in Fourier domain
    \begin{align}
        Y[k] = X[k]\cdot D[k] = a[t]*S[t]D[k]W_{K}^{\tau} 
    \end{align}
    The final step in computing the CAF is taking the inverse FFT
    \begin{align}
        R_{xd}(f_{D},\tau) = \textsc{IFFT}_{K} \{Y[k]\} = a[k]\sum_{n=0}^{K-1} s[n]d[k-n] 
    \end{align}
    The ``peak metric'' for a given local replica is found by maximizing the squared magnitude of the correlation grid. At the peak where the signal component $s[k]$ and the local replica are identical, this ideally reduces to
    \begin{align}
        S_{\textsc{max}} = |R_{xd}(f_{D},\tau)|^{2}\big|_{f_D\approx\hat{f}_D,\tau\approx\hat{\tau}} = |a|^2 |K|^2
    \end{align}
    where $S_{\textsc{max}}$ is the maximum peak and $R_{xd}(f_{D},\tau)$ is the search grid.
    Rearranging this we find an expression for the amplitude of the input signal in terms of the peak metric
    \begin{align}
    |a| = \frac{\sqrt{S_{\textsc{max}}}}{K}
    \end{align}

\noindent Equipped with all the above information, the recovery signal is generated.
LSR performs this iterative cancellation process for all the satellites.\bigskip

\noindent \emph{\pseurecuc{}:}
Specific attack scenarios result in tracking failure \ie the receiver is unable to extract the navigation message. 
Such a scenario is possible when an adversary introduces extreme interference that buries the legitimate signals under the noise floor. 
As a result, the adversary will corrupt or distort navigation bits of the legitimate signal. 
Other kinds of interference may result in a situation in which the correlation peaks being close enough together that the adversary flips the bits in the legitimate navigation message. 
Even if \recvname can recover the legitimate peak, it won't be able to successfully track and decode navigation bits. 
This leads to incorrect calculation of location. 
The pseudorange rectifier enables \recvname to correct these ambiguities and aid in the recovery of location.
An important assumption is that the attacker manipulates the location by changing the arrival time of the signals and keeps the navigation messages the same \ie legitimate and adversarial messages are the same.

Commercial GPS receivers use a common reception time technique \cite{rao2012can} to calculate pseudorange to the satellite, an essential component in PVT calculation.
In this technique, a common reception time, which is usually 65-85 ms~\cite{rao2012can}, is set across all the channels as the propagation time of the closest satellite's signal.
The receiver calculates the propagation time of signals from other satellites relative to this reference. 
Modern GPS receivers maintain a sample counter for accurate time measurement.
According to this technique, pseudorange is calculated as follows:
\begin{equation}
    P^{i} = c (t_{ref} + t_{rx} + \tau^{i})
\end{equation}
\noindent where $P^{i}$ is the pseudorange measurement for $i^{th}$ satellite, $c$ is the speed of light, $t_{ref}$ is the initial reference time (usually 65-85 ms \cite{rao2012can}), $t_{rx}$ is the receiver time maintained by a sample counter, and $\tau^{i}$ is the code phase delay of $i^{th}$ satellite.\bigskip

\noindent \recvname attenuates the adversarial peak and obtains tracking parameters of the legitimate peak. 
However, it doesn't track the legitimate peak.
Instead, it starts tracking the adversarial peak and obtains adversarial navigation messages. 
A stealthy attacker will keep navigation messages the same and change only the signals' ToA. 
It offsets the sample counters by $\tau_{at}^{i} - \tau_{l}^{i}$ where $\tau_{at}^{i}$ is the code phase delay of $i^{th}$ satellite of the attacker and $\tau_{l}^{i}$ is the code phase delay of $i^{th}$ legitimate satellite obtained during the peak recovery. 

\begin{equation}
    P^{i}_{l} = c (t_{ref} + t_{rx} + \tau^{i}_{at} -  \Delta\tau^{i})
\end{equation}

\begin{equation}
    \Delta\tau^{i} = \tau^{i}_{at} - \tau^{i}_{l}
\end{equation}

\noindent Substituting (13) in (12) we get (11).
In this way, \recvname can obtain legitimate pseudoranges ($P^{i}_{l}$) by rectifying ToA of adversarial signals. 

\label{sec:lsr}
\section{Implementation}
\label{sec:implementation}
\begin{figure}[t]
   \centering
   \includegraphics[width=0.9\columnwidth]{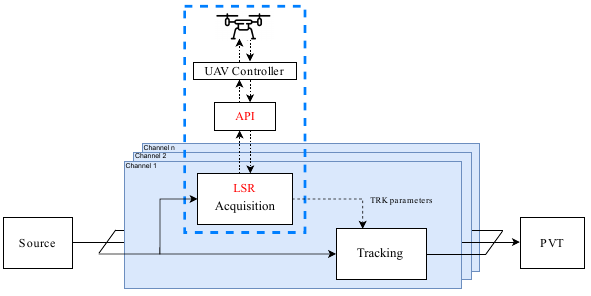}
   \caption{LSR is implemented as part of GNSS-SDR and API is implemented as part of the UAV flight controller.}%Schematic showing the implementation of LSR as part of GNSS-SDR and API as part of the UAV flight controller.}  
   \label{fig:gnss-sdr-impl}
\end{figure}
The two sub-systems which make up SemperFi are implemented independent of one another:
the API is implemented at the flight controller level while the LSR alongwith spoofing detector is implemented in GNSS-SDR as part of the acquisition block.
These two components interact with each other over a TCP socket.
We implemented LSR module of \recvname{} in GNSS-SDR~\cite{fernandez2011gnss} an open-source software-defined GNSS receiver written in C++. 
We implemented the API module using consumer drones.
Refer to~\Cref{fig:gnss-sdr-impl} for a schematic of the implementation.
GNSS-SDR follows GNURadio architecture and supports the processing of pre-recorded signals from a file source and software-defined radio frontends like a USRP~\cite{ettus}.
GNSS-SDR follows a hardware receiver's design as described in~\Cref{sec:background}, except all the components are implemented in software. 
Signals from individual satellites are processed by individual \emph{channels}. 
Each channel is like a hardware pipeline of various GPS signal processing blocks, including acquisition, tracking, and PVT calculation.
At run-time, the GNSS-SDR builds the receiver using these blocks based on specifications from a user-defined configuration file.
This allows loosely coupled operations.
In our implementation and evaluation, we use software-defined radio hardware platforms manufactured by Ettus Research~\cite{ettus}, specifically, USRP B210 and N210 with SBX-40 daughterboard, for recording and providing raw data. 

\begin{figure}[t]
    \centering
    \includegraphics[width=0.9\columnwidth]{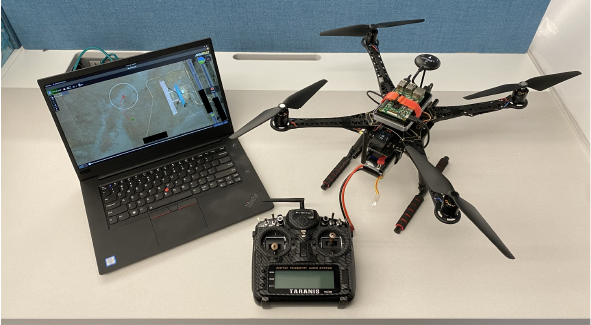}
    \caption{Hardware setup showcasing the Holybro S500 drone, the radio controller and the ground control station.}  
    \label{fig:s500}
 \end{figure}
\paragraph{\apiuc (API)}
In \recvname, API is implemented as an independent module that interacts with the LSR
This was implemented on an unmanned aerial vehicle in a simulated environment as well as on a DJI Flamewheel F450 and a Holybro S500.
These drones were specifically chosen as they support Pixhawk 4~\cite{pixhawk}, an advanced autopilot system and ArduCopter~\cite{ardupilot} firmware. 
Refer to~\Cref{fig:s500} for the hardware setup.
A spoofing attack can cause errors in EKF estimates and raise EKF variance errors as GPS and EKF estimates do not match. In such a case, ArduCopter activates EKF failsafe.
Moreover, ArduCopter may raise GPS glitch errors and activates the GPS failsafe. 
By default, ArduCopter switches to \emph{LAND} mode and lands at the current location.
To prevent this, we temporarily disabled EKF and GPS failsafe by manipulating the \emph{FS\_OPTIOINS} parameter.
The pseudorandom maneuver is implemented as a sequence of left/right turns determined at run-time.
The algorithm generates a pseudorandom set of velocity vectors and instructs the autopilot to fly that heading for a specified time.
To achieve this, we used the \emph{SET\_POSITION\_TARGET\_LOCAL\_NED} MAVLink message type to instruct the drone to move with a specified velocity and for a specific time.
In our implementation, we used \emph{DroneKit}~\cite{dronekit} installed on a Raspberry Pi 3B+ to generate the maneuver and instruct the flight controller to execute it.
A specific sequence of these messages then carries out the entire maneuver.
Once the UAV completes the maneuver, API performs the correlation operation as described in~\Cref{sec:api} and notifies LSR over a TCP socket.

\paragraph{\rssuc{} (LSR)}
In SemperFi, we implement LSR as a part of GNSS-SDR's acquisition module.
As mentioned earlier, we use auxiliary peak based spoofing detection technique as proposed in~\cite{ranganathan2016spree}. 
We modified the acquisition block such that spoofing detection is triggered every time the acquisition block is activated.
This allows \recvname{} to recover from hard spoofing attacks that result in a loss of lock.
Positive detection of an adversarial signal triggers further processing that includes peak identification, recovery signal generation, and signal recovery.
GNSS-SDR allows external communications using TCP sockets as outlined here~\cite{fernandez2012open}.
This enables GNSS-SDR to interact with the UAV's flight controller responsible for performing peak identification maneuvers.
Once the API validates spoofing and provides peak information, LSR proceeds to the cancellation and recovery state.
At this stage, LSR has the peak information and a rough estimate of the Doppler and the code phase delay of the satellite signal.
The accuracy of parameter estimates is directly related to the degree to which the adversarial peaks may be attenuated; \recvname performs re-acquisition using a more refined grid search to obtain more precise estimates.
After performing a narrow search, LSR generates a replica of the satellite signal using the tracking parameters estimated in the two-step acquisition process.
LSR also estimates the satellite signal's amplitude using the method described in~\Cref{sec:lsr}.
We use the Vector-Optimized Library of Kernels~\cite{west2016vector} function to perform vector operations.
These functions provide a significant boost to performance and reduce computation time.
Once the signal is regenerated, it undergoes phase correction and cycles through phase shifts to determine maximum attenuation.
Due to inaccuracies in the amplitude, Doppler, and the code phase delay estimates, a single attempt at recovery will not wholly attenuate the adversarial peak.
SemperFi iterates the entire acquisition and recovery process until the legitimate signal is stronger than the adversarial signal.

\noindent \emph{\pseurecuc{}:}
This module is implemented as an optional component in the tracking module and is disabled by default. 
The receiver enables \pseurec{} if the navigation message decoder fails to detect a preamble even after tracking the correct peak.
Even if the navigation message decoder can find preamble and decode the navigation bits, there is a possibility that adversarial peak interferes with correct PVT estimation. 
In these cases, the receiver will activate \pseurec{}. 
\pseurecuc{} can also be activated manually by setting a flag in the receiver configuration file. 
When \pseurec{} is activated, the tracking module tracks the adversarial peak instead of the legitimate peak. 
It, however, still obtains tracking parameters of the legitimate peak. 
It uses legitimate and adversarial code phase information to calculate $\Delta\tau^{i}$. 
Code phase information and subframe start pointer determined by preamble position in a buffer of samples are used to determine the ToA of satellite signals. 
A sample counter accurately maintains this information. 
$\Delta\tau^{i}$ is used to offset sample counters appropriately. 
The receiver still decodes adversarial navigation messages; however, it uses the ToA of legitimate signals for pseudorange calculation to calculate the correct PVT solution.
\section{Security and Performance Evaluation}

We evaluate \recvname{} and showcase its performance in recovering legitimate GPS signals under various attack settings and signal traces. 
Specifically, we use three different datasets that contain both spoofing and legitimate signals: i) Synthetic GPS signals generated using COTS GPS simulators, ii) a public repository of GPS spoofing signals (TEXBAT)~\cite{humphreys2012texas}, and iii) recorded real-world GPS signals. \bigskip

\subsection{Evaluation Traces}
\label{sec:sig_traces}
\textbf{GPS Simulator:} We performed most of our evaluation on synthetic signal traces generated locally using GPS-SDR-SIM\cite{osqzss2015gpssim}, an open-source tool for generating GPS signals. 
This provides granular control over signal properties such as power levels, temporal delays, and Doppler shifts; thus enabling the user to generate a variety of spoofing scenarios.
We evaluated \recvname against both static (stationary locations) and dynamic scenarios (motion trajectories).

These signals were transmitted using two USRP B210s, one each for the legitimate and attacker signal.
We recorded the signals using a USRP N210 at a rate of $10\unit{MSa/s}$. 
We wired all RF-frontends to prevent signal leakage as it is illegal and hazardous to transmit GPS signals..
For static and dynamic scenarios, we picked locations in downtown San Francisco.
We generated the attacker's signal such that the obtained location is at a specific offset from the legitimate location. 
We picked locations with the offset increasing in steps of 500 m up to a maximum spoofed offset of 3500 m.
The offset locations were specifically selected to simulate various scenarios.

\noindent \textbf{Texas Spoofing Test Battery (TEXBAT):} TEXBAT is a set of civilian GPS spoofing scenarios that are a standard for evaluating spoofing countermeasures. 
The repository consists of spoofing signals traces that include both position and time push scenarios.
TEXBAT also provides scenarios where the attacker's signals and the legitimate signals are synchronized, similar to the strong seamless-takeover attack. 
We evaluate the effectiveness of \recvname against both static and dynamic position push.
These signal traces were recorded at $25\unit{MSa/s}$.
The traces are 7 mins long, and the attacker starts spoofing roughly $90-100\unit{s}$ into the signal trace.

\begin{figure}[t]
    \centering
    \includegraphics[width=0.8 \columnwidth]{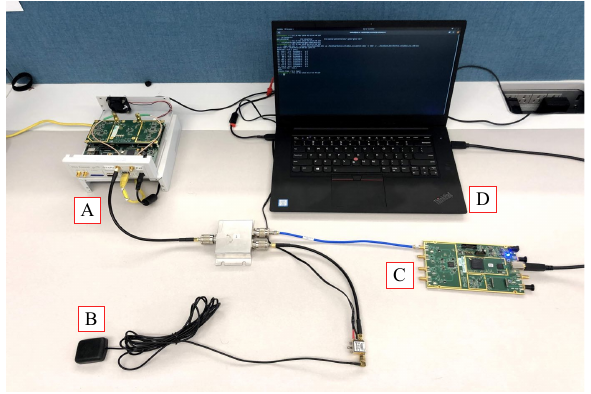}
    \caption{Signal recording setup A) GPS signal RX (USRP N210), B) ANT-555 active GPS antenna with a 5V bias-tee, C) GPS signal TX and D) GPS simulator control unit.}  
    \label{fig:live_setup}
\end{figure}

\noindent \textbf{Live GPS Recordings:} We also evaluated \recvname against a combination of live legitimate GPS signals and attacker signals. 
This scenario covers the real-world spoofing scenario where the attacker transmits spoofing signals while the receiver is locked on to legitimate signals. 
We recorded a set of real-world GPS signal traces through extensive war-driving in our locality\footnote{location anonymised.}.
We recorded the legitimate GPS signals using the setup shown in~\Cref{fig:live_setup}. 
We captured the GPS signals using an ANT-555 antenna supplied with a 5 V DC power supply.
We combined the received signal with the attacker signals using a combiner. 
We used GPS-SDR-SIM to generate attacker's signals. 
The spoofed location was set to 4.1\unit{km} away from the original location. 
Hard-wiring the attacker allowed us to test in a best-case scenario for the attacker as they have a clear channel to the victim receiver and evaluate its performance in eliminating the spoofing signal.

\subsection{Evaluation Metrics and Results}
In this section we evaluate our implementation of \recvname{} and its components.
We evaluate API's performance by studying the maneuvers' feasibility and the ability to correctly distinguish legitimate signals and adversarial signals.
The evaluation was performed using real drones and also using Gazebo~\cite{gazebo}, a robotics simulator.
Evaluation metrics for evaluating the recovery process are amplitude estimation accuracy, the accuracy of the recovered location, and the time required to perform recovery.
To further evaluate the results, we study how attacker synchronization and the power advantage an attacker has over the legitimate signals affect the recovered location's accuracy.
We also evaluate the effects of jamming attacks on the drones.

\paragraph{Adversarial Peak Identification:}
\begin{figure}[t]
    \centering
    \includegraphics[width=\columnwidth]{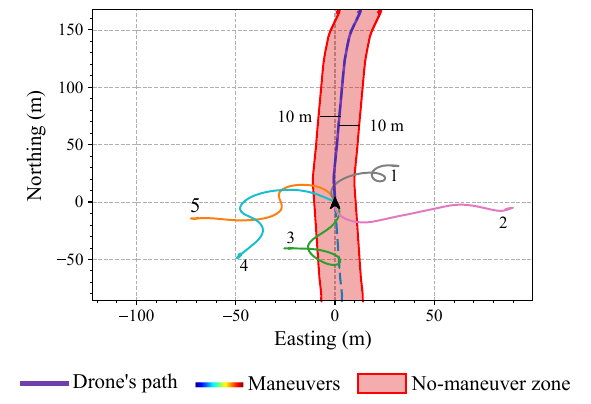}
    \caption{Pseudorandom trajectories generated by API. The maneuvers are triggered at location (0,0). The region marked in red is considered to be the ``No-maneuver'' zone as the attacker has higher probability of estimating the drone's movement.}  
    \label{fig:api_all_maneuvers}
\end{figure}
It was evaluated in a simulated environment using Gazebo~\cite{gazebo} and on real drones.
For evaluation, we follow a scenario where the attacker is successful in executing a spoofing attack.
The attacker executes either a seamless takeover or a hard spoofing attack.
In either case, API will trigger the maneuver when the peak separation is more than 500\unit{ns} as explained in~\cite{ranganathan2016spree}. This is, however, configurable.
We evaluate the peak identification strategy by studying the feasibility of the maneuvers and the ability to accurately distinguish between the trajectory of the drone as tracked by IMU sensors and GPS receiver. 
IMU sensors tend to accumulate errors over time. 
For a commercial navigation grade MEMS sensor, these errors or `random walks', can be up to 1.59\unit{km/hr}\cite{narain2019security}. 
Thus for a 60\unit{s} maneuver, position estimates can drift up to 26\unit{m}.
Based on these characteristics, the API uses a threshold of 5\unit{m} to decide whether the receiver is tracking legitimate signals or adversarial signals.
Researchers in \cite{woodman2007introduction} show that a sensor fusion algorithm that combines IMU measurements with a magnetometer can significantly increase the sensors' accuracy.
\begin{figure}[t]
    \centering
    \includegraphics[width= 0.9 \columnwidth]{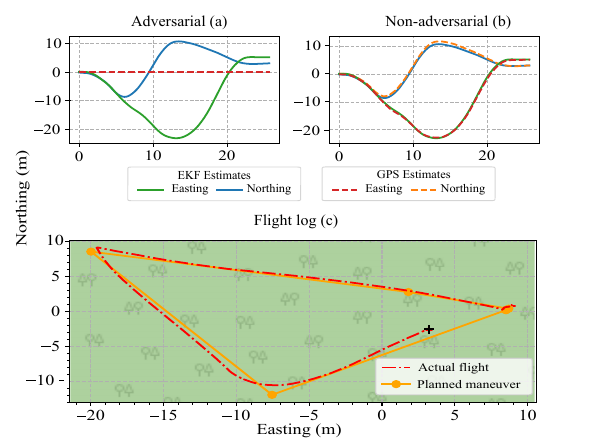}
    \caption{(a) shows the comparison of position estimates of EKF and GPS in an adversarial setting where the attacker is unable to estimate the drones maneuver. (b) shows a comparison of position estimates of EKF and GPS in a non-adversarial setting. The trajectories in (b) show trivial deviation whereas, trajectories in (a) show significant deviation. (c) shows the flight log, the UAV's actual trajectory and the maneuver generated by the API.}
    \label{fig:real_maneuver}
\end{figure}
We tested maneuvers with flight time of 10\unit{s} to 60\unit{s}.
After a thorough evaluation, we concluded that increasing the time duration will: i) cause more inaccuracies because of IMU drifts and ii) allows the attacker an increased window to predict the maneuver and spoof.
To accurately determine malicious signals, it is essential to follow a trajectory that is unknown to the attacker. 
Performing the maneuver using the current trajectory will indeed reduce the required time.
However, a major limitation is that the current trajectory is either defined by the attacker or known to the attacker.
A random trajectory increases complexity and time but, since the attacker is unaware of this trajectory, it has a better chance of succeeding. 
\Cref{fig:api_all_maneuvers} shows few pseudorandom trajectories generated by API.
Based on factors like accuracy of civilian GPS, IMU errors and drifts, and the EKF leash described in~\cite{noh2019tractor} we determined a ``No-maneuver'' zone that is 20 m wide.
It is necessary for the drone to execute a maneuver that deviates from the current trajectory and one that is not straight.
Trajectory no. 2 in~\Cref{fig:api_all_maneuvers} deviates from the current trajectory, however, follows a straight line and hence it is possible for an attacker to predict it and generate spoofing signals accordingly.
Thus, it is necessary to perform this invasive maneuver.
Flight time for these maneuvers was between 10-20\unit{s}.
For deciding the maneuver, it is vital to consider the drifts that IMUs will experience, the speed at which the UAV can execute maneuvers (UAVs will reduce speed significantly when turning), and the accuracy of GPS measurements. 
The surrounding terrain and weather play an important role in determining the maneuver.
We assume that if the drone is operable for our evaluation, it can perform the required maneuvers.
ArduCopter can provide accurate positioning with IMU measurements for 10\unit{s}~\cite{arducoptergps}, after which IMU measurements start drifting due to lack of rectification. 
For successful identification, UAV should complete the maneuver under 20\unit{s} and travel at-least 30\unit{m} from the original position where the identification maneuver was triggered. 
~\Cref{fig:real_maneuver}(a) and (b) shows the deviations in trajectories as estimated by EKF and GPS sensors in adversarial and non-adversarial settings, respectively.
The correlation algorithm explained in~\Cref{sec:api} can detect such deviations.
~\Cref{fig:real_maneuver}(c) shows the actual flight path flown by the UAV and the project path as generated by the maneuver generation algorithm.\footnote{The actual maneuver takes 15\unit{s}. The plot includes the time the drone takes to arm/disarm.}
We used a Holybro S500 drone to generate the data in~\Cref{fig:real_maneuver}.
Prior works like~\cite{savior2020usenix} show the use of sensor fusion to detect spoofing attacks.
However, these solutions may not provide reliable attack detection as an attacker can stealthly introduce spoofing signals.
\begin{figure}[t]
    \centering
    \includegraphics[width=\columnwidth]{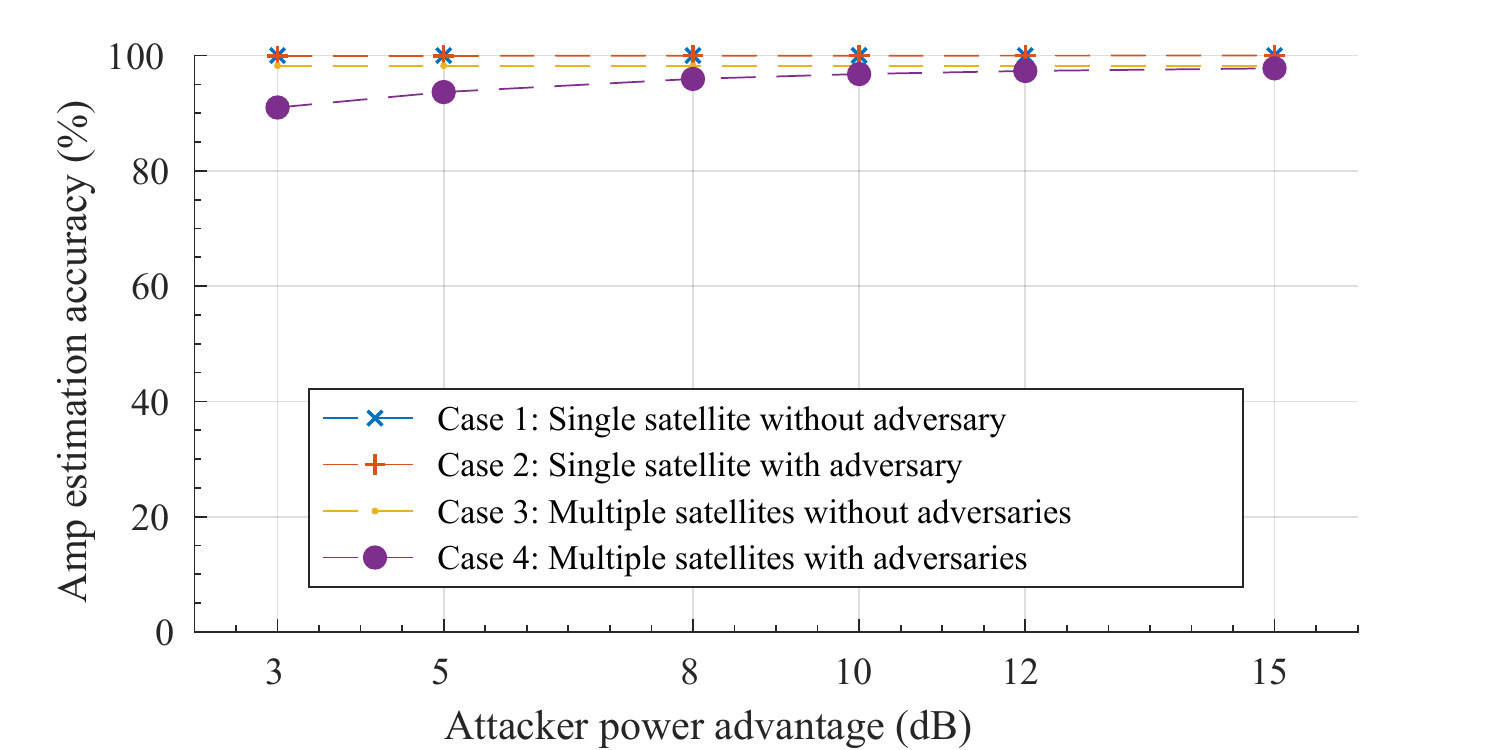}
    \caption{Amplitude estimation accuracy in various scenarios. Attacker power advantage does not apply to cases 1 and 3. In case 4, each satellite is spoofed. Power advantage refers to the advantage that the attacker's signal has over the legitimate signal.}
    \label{fig:amp_eval}
\end{figure}

\paragraph{Amplitude Estimation:}
It plays a vital role in successful signal recovery. 
In SemperFi, we leverage the max CAF value or the correlation coefficient value to estimate the original signal's amplitude. 
In this strategy, the estimate's accuracy is susceptible to various factors like interference caused by signals from other satellites, the presence of adversarial signals, and artifacts introduced by a wireless channel. 
For evaluating the accuracy, we conducted an experiment where we executed amplitude estimation in four cases. Refer to~\Cref{fig:amp_eval} for results.  
The accuracy of amplitude increases as the attacker's power advantage increases.
As a result of inaccuracies in amplitude estimation caused by Doppler shifts, clock skews, and phase shifts, SemperFi may perform multiple iterations to attenuate the adversarial signal and recover successfully.

\paragraph{Recovered Location Accuracy:}
We evaluate \recvname's effectiveness in eliminating the spoofing signal by determining the location's accuracy after passing through the various blocks of \recvname. 
We use the Universal Transverse Mercator (UTM)~\cite{usgsutm} system to present our location accuracy results.
We evaluated the performance of \recvname against both static and dynamic scenario spoofing attacks present in the datasets described in~\Cref{sec:sig_traces}.

\begin{figure}[t]
    \centering
    \includegraphics[width=0.8 \columnwidth]{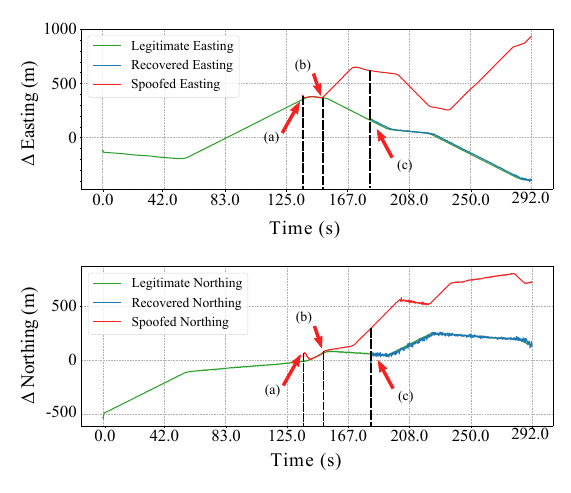}
    \caption{Changes in recovered path and legitimate path coordinates as represented in UTM coordinate system for recovery of dynamic spoofing scenario (GPS Simulator).}  
    \label{fig:recovery_utm_plot}
\end{figure}

\begin{figure}[t]
    \centering
    \includegraphics[width= 0.8 \columnwidth]{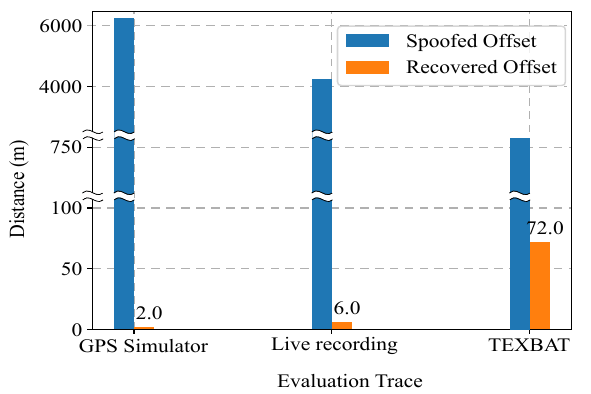}
    \caption{The recovered offset and spoofed offset for three scenarios.}
    \label{fig:static_recovery}
\end{figure}

\begin{figure}[t]
    \centering
    \includegraphics[width= \columnwidth]{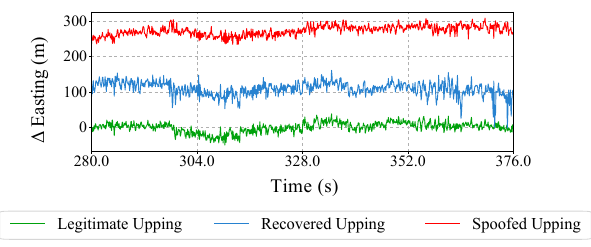}
    \caption{Variations in recovered path and legitimate path coordinates represented in UTM coordinate system.}  
    \label{fig:recovery_ds6_plot}
\end{figure}

First, we evaluate the performance of \recvname against the dataset generated using GPS signal generators. 
The UTM plots depict the variations in locations and a timeline of events.
As seen in~\Cref{fig:recovery_utm_plot}, at point (a) the spoofer starts. 
Roughly for 15 s, the attacker is in sync with the legitimate receiver. 
During this time, the acquisition plot shows no auxiliary peaks. 
After 15 s at point (b), the attacker starts introducing offsets in the calculated location. 
As a result, the receiver starts deviating from the expected trajectory. As soon as peak separation is enough to rule out multi-path transmissions, \recvname{} is activated, and at point (c), the receiver starts following the recovered trajectory in-spite of adversarial presence. 
Average deviation of recovered path from legitimate path is 10.1\unit{m} (Easting) and 2.6\unit{m} (Northing).
\Cref{fig:static_recovery} shows the recovery operation results on static scenarios across all three datasets. 
GPS simulator traces where the spoofed offset is 6.2\unit{km} with recovered offset of 2\unit{m}. 
Live recording with a recovered offset of 6\unit{m}. 
TEXBAT's power matched position push scenario where the attacker spoofs only in $Z$ plane.
Figure~\ref{fig:recovery_ds6_plot} shows the performance of \recvname against TEXBAT's dynamic position push. 
Note that TEXBAT's position push consists only of altitude, which is known to be error-prone for GPS~\cite{rothacher2002estimation}. 
\recvname was able to recover the spoofed $Z$ plane offset with an accuracy of 108\unit{m}.


\begin{figure}[t]
    \centering
    \includegraphics[width=0.8 \columnwidth]{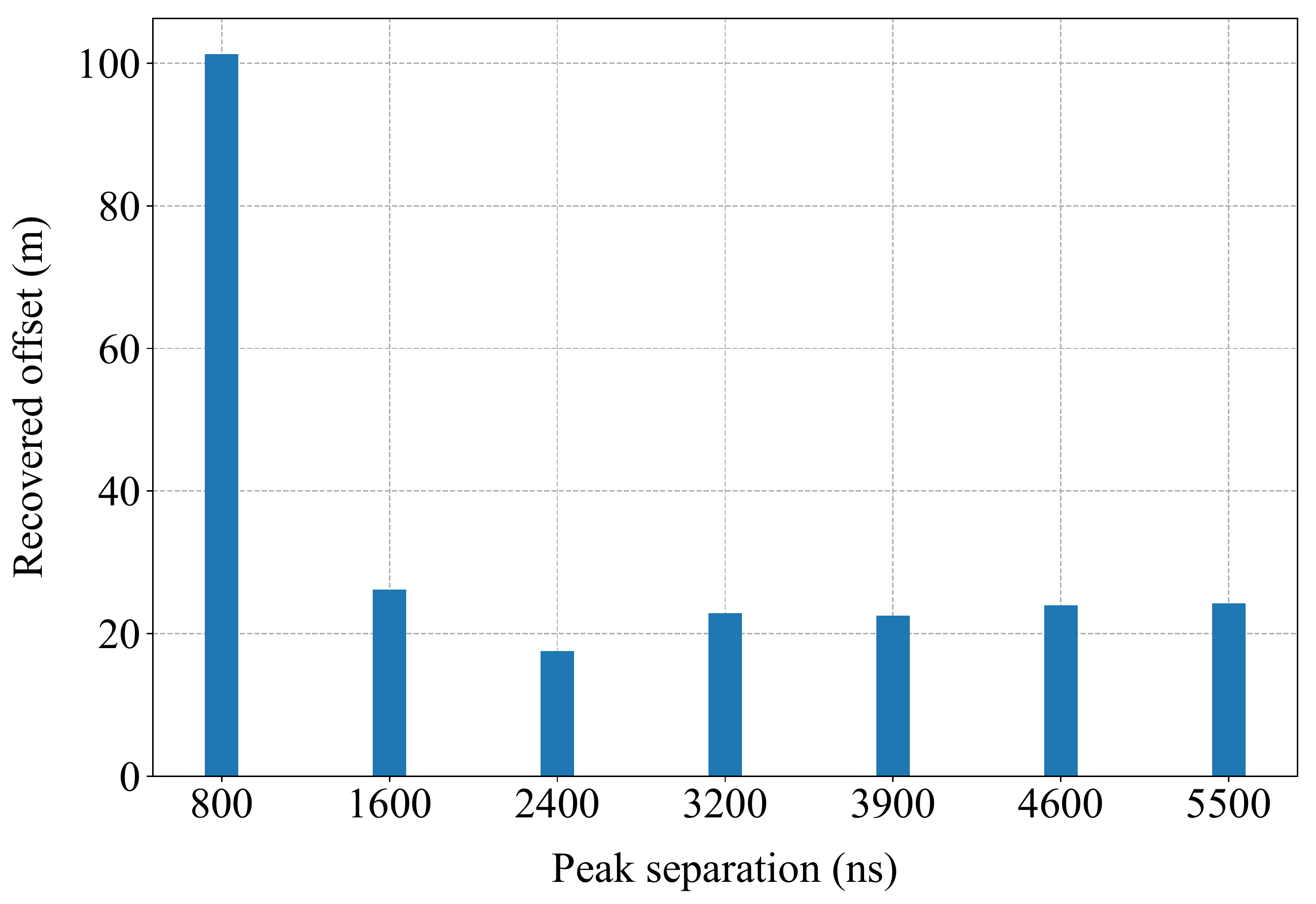}
    \caption{The effect of peak separation on accuracy of the recovered location. The closer the peaks, the harder it gets to accurately track them.} %with and without \pseurec}
    \label{fig:ase_pse}
\end{figure}

\begin{figure*}[t]
    \centering
    \includegraphics[width=0.8 \textwidth]{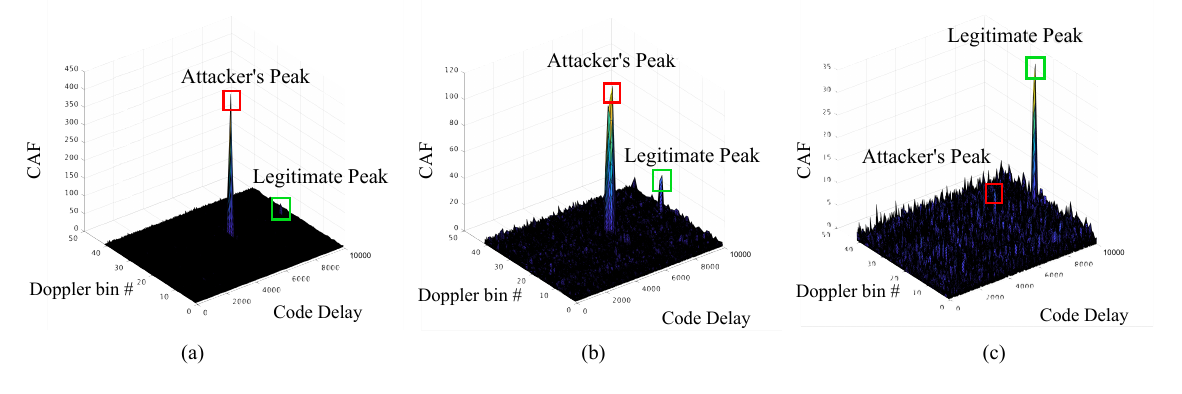}
	\caption{Two step signal attenuation of a strong adversarial signal. (a) shows the original acquisition plot, (b) shows acquisition plot where legitimate peak is slightly visible and (c) final acquisition plot with fully suppressed adversarial peak.}
	\label{fig:high_power_acqplot}

\end{figure*}

\begin{figure}[t]
    \centering
    \includegraphics[width=0.8\columnwidth]{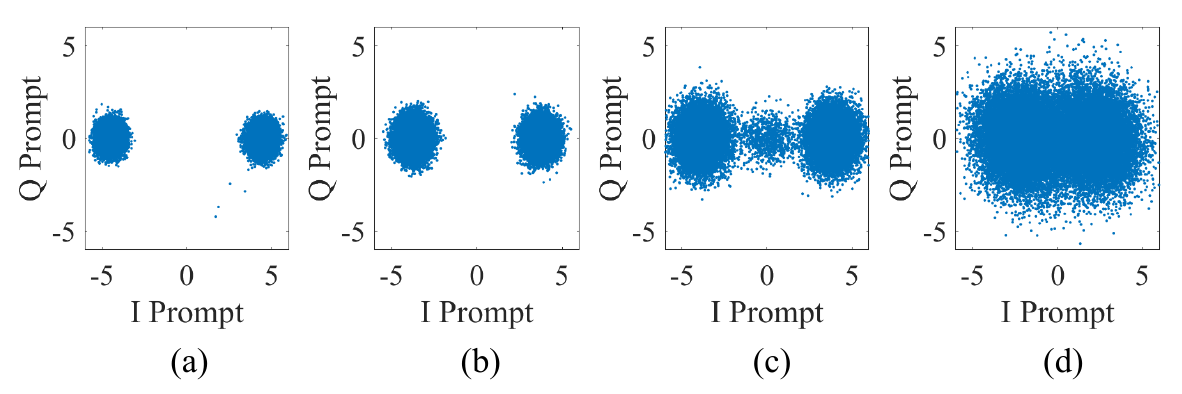}
    \caption{Discrete time scatter plot of recovered nav message where attacker has (a) 3 dB , (b) 5 dB, (c) 10 dB  and (d) 15 dB power advantage. A powerful attacker adds noise and hence distorts legitimate nav messages.}
    \label{fig:3db_15db_iq}
\end{figure}

\begin{figure}[t]
    \centering
    \includegraphics[width= 0.9 \columnwidth]{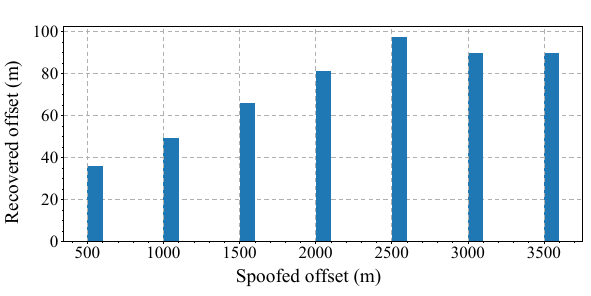}
    \caption{Spoofed offset vs offset in recovered location for attacker with 15 dB power advantage. \recvname uses \pseurecuc{} for recovery. For locations refer to~\Cref{sec:sig_traces}.}
    \label{fig:pse_15db}
\end{figure}

\paragraph{Attacker synchronization:}
One major factor which affects recovered locations is attacker synchronization with legitimate signals. 
In other words, the effectiveness of eliminating spoofing signals depends on the temporal shifts in the ToA of legitimate and spoofing satellite navigation messages. 
The closer the synchronization, the harder it is to recover entirely without additional processing. 
We evaluated the effects of attacker synchronization by generating spoofing scenarios where the attacker spoofs locations with an offset in the increments of 500 m from the original position.
This results in a corresponding temporal shift between the attacker's spoofing signal and the legitimate signal. 
The minimum peak separation was 800 ns at 500 m, and the maximum peak separation 5500 ns at 3500 m. 
Note that this peak separation depends on the satellite constellation at any point in time. 
~\Cref{fig:ase_pse} shows the results of this experiment.
Evenly spaced separation in the distance (m) is replaced by corresponding mean peak separation in tracked satellites' represented in nanoseconds. 
Peak separation directly affects how the attacker's signals interact with legitimate signals ( \eg 800 ns replace 500 m).
This poses a challenge to the tracking loops, and as a result, the tracking loops undergo signal cross-over.
I.e., the peaks are so close that the tracking loop starts tracking the wrong signal. 
This is evident from higher recovered location variation for scenarios with closer peaks.

\paragraph{Effect of Attacker's Power Advantage:}

Finally, we evaluate the performance of \recvname against attackers with varying power levels up to 15 dB. 
Note that in seamless takeover attacks, the maximum power difference required to execute the attack successfully is not more than $2-3\unit{dB}$~\cite{tippenhauer2011requirements,humphreys2012texas}. 
TEXBAT repository's seamless takeover attack data-trace has a power difference of not more than 10\unit{dB}.
We created spoofing scenarios where the attacker has a power advantage of 3 to 15 dB. 
\recvname can attenuate stronger peaks and make the suppressed weaker legitimate peaks visible in the acquisition plot.
~\Cref{fig:high_power_acqplot} shows a multi-stage attenuation process for an adversary with 15 dB power advantage. 
However, as seen in discrete-time scatter plot in~\Cref{fig:3db_15db_iq}(d), in the case of an attacker with a 15 dB power advantage, the adversarial signal introduces much noise, which distorts the navigation bits. 
In such a scenario, despite the reduced accuracy the receiver can switch to pseudorange rectifier and recover the correct location by rectifying pseudoranges.
~\Cref{fig:pse_15db} shows the results of signal recovery in the presence of an attacker with a 15 dB power advantage.

\begin{table}[t]
    \centering
    \begin{tabular}{|c|c|}
        \hline
        \textbf{Model} & \textbf{Processing time} \\ \hline
        RPi 3B+        & 11.57 s/itr              \\ \hline
        RPi 4          & 6.98 s/itr               \\ \hline
        Jetson Nano    & 8.35 s/itr               \\ \hline
        Jetson Xavier  & 5.19 s/itr               \\ \hline
        Intel Core i7  & 2.79 s/itr               \\ \hline
        Intel Xeon     & 2.12 s/itr               \\ \hline
        \end{tabular}
    
    \caption{A comparison of time required by the corresponding system to perform one iteration of signal cancellation.}
    \label{tbl:performance}
\end{table}

\paragraph{Real-time performance:}

Implementation of SemperFi on GNSS-SDR allows us to deploy SemperFi on various mobile platforms and embedded systems. We evaluate the performance by deploying and executing SemperFi on the following systems. i) Raspberry Pi 3B+, ii) Raspberry Pi 4, iii) NVIDIA Jetson Nano, iv) NVIDIA Jetson Xavier, v) Intel Core i7\footnote{https://www.dji.com/manifold-2}, and vi) Intel Xeon E5-2630\footnote{not currently used in any UAV platform and ported only for comparison}.
These systems are some of the standard systems used as flight controllers onboard UAVs.
We use signal traces as described in~\Cref{sec:sig_traces} for evaluating the performance. The sampling rate of $10\unit{MSa/s}$ plays a significant role in determining the performance of SemperFi as it is directly related to the processing overhead.
Our primary evaluation metric is the time required per iteration of cancellation.
It is important to note that GNSS-SDR is itself a resource-demanding application. 
Refer to~\Cref{tbl:performance} for a comparison showing each system's performance. 
Each platform may take up to 5 iterations for complete recovery depending on the accuracy of the estimates.
Thus, complete signal recovery may add delay to the calculation of the PVT solution; in the case of Jetson Xavier, for example, by 25.96\unit{s}.
It is important to note that these values are from a sub-optimized version of SemperFi. 
It is possible to improve the performance by optimizing SemperFi for a specific system that leverages its unique characteristics and features. For example, SemperFi can be re-programmed to use CUDA cores available of NVIDIA Jetson Nano and Xavier.
In general, it is best to deploy SemperFi on an FPGA as it will significantly improve the performance.

\paragraph{UAVs' Resilience to Jamming Attacks:}
Unlike in the spoofing attack, the attacker's primary goal is to cause a denial of service attack on GPS in jamming attacks.
Hard spoofing attacks as described in~\cite{noh2019tractor} and jamming attacks will cause the receiver to lose the lock.
However, in case of a hard spoofing attack, the receiver will be able to re-acquire the lock, while on the other hand, in case of a jamming attack, it will not.
For a successful jamming attack, an attacker has to transmit noise, or a simple amplitude modulated continuous wave at the GPS carrier frequency.
An attacker can also transmit modulated dummy GPS navigation messages (random 1s and 0s) using GPS modulation techniques and execute a successful denial of service attack,
It is important to note that GPS receivers provide a processing gain of 43.1\unit{dB}.
Thus, the jamming signal should be strong enough to overpower the legitimate signal at the receiver.
Since we performed GPS jamming experiments in an anechoic chamber, creating a GPS denied region was accomplished by simply turning off our GPS transmitter.
All modern drones are equipped with fail-safe mechanisms. 
The drone activates these fail-safes in case of GPS is unavailable.
These fail-safes can be: i) Land at the current location, ii) Return to home using IMU sensors, or iii) hover in the current location until a GPS fix is re-established.
In our case, the drone was programmed to land at the current location. 
Prior work ~\cite{olsen2003jamming} has showcased various GPS jamming techniques and provides a comprehensive analysis of GPS receivers' susceptibility to jamming.
\section{Discussion}
\paragraph{Flexible design:} SemperFi is designed to be flexible and versatile.
In addition to integrating SemperFi into the acquisition module as shown in~\Cref{sec:implementation}, we can use \recvname as a plug-in module that can filter out adversarial signals and pass on legitimate signals to a conventional receiver.
This mode of operation requires minimal modifications to the existing receivers.
SIC technique allows \recvname to be used as a plug-in module.
Furthermore, SemperFi's capabilities can also be extended to other satellite navigation systems like GALILEO as they follow a similar operating principle of code division multiple access using spreading codes and computation of pseudoranges.
As a result of lack of robust countermeasures, they face similar security issues.

\paragraph{Limitations:}
The current design restricts \recvname to aerial vehicles. 
It is challenging to design these maneuvers for terrestrial vehicles due to the vehicle's mobility constraints. 
As shown in ~\cite{narain2019security}, an attacker can exploit the short-term stability of IMU sensors due to predictable maneuvers and limited mobility. 
As seen in recent GPS-related incidents, ~\cite{tesla2019hack,shanghai2019gpshack} road and oceanic navigation pose a challenge in inefficient peak identification given the limitation of performing maneuvers.
Even if the drone is operable, frequently changing weather conditions, especially wind vectors, can affect the drone's maneuverability, especially high-velocity crosswinds.
However, the algorithm can be modified to work with crosswinds by determining the force and velocity of wind as proposed in~\cite{neumann2015real} or by equipping the drone with solid-state anemometers.
An attacker aware of SemperFi can indeed start transmitting different spoofing signals periodically. 
However, this will result in an attack similar to jamming, which can be trivially detected. For an attacker to stealthily circumvent SemperFi, the spoofing signals should reflect UAV's random maneuver to match the GPS derived motion with the IMU derived one. An attacker capable of generating spoofing signals in real-time based on the random maneuver will succeed, and the feasibility of executing such an attack remains to be studied. 
An attacker will have to predict the maneuver accurately to be able to evade peak identification.
To do this, an attacker will have to track the drone in real-time with delays of only a few milliseconds as in this time; the attacker has to predict the next coordinate to spoof, construct the GPS signal and transmit. 
The refresh rate of typical GPS receivers provides the attacker about 200\unit{ms} to perform all these tasks. 
An attacker can do this by deploying passive quadcopter detection RADARs as described here~\cite{guvencc2017detection, DroneDefense, fang2018experimental}.
The above references all detect and track drones using advanced techniques like ultra-wideband scanning. 
However, none generate GPS spoofing signals to emulate the drone's behavior.
Given the time constraints, it is hard for an attacker to generate spoofing signals and track the drone's movements.
Concerning the generalization of the proposed solution, Semper-Fi exploits the short-term stability of IMU and executes quick maneuvers. 
Designing these maneuvers for terrestrial vehicles is a challenge due to the constraints on the vehicle's mobility. 
The problem is similar when a UAV is flying between obstacles. Note that the attacker also has similar constraints to force the drone on a different path successfully. 
A potential solution is to let the UAV decide the maneuver in real-time based on its degree of freedom; however, this might increase the probability of the attacker guessing the maneuver correctly. %\todo{Expand the above and integrate it with your writing.

Another limitation of \recvname is that tracking legitimate signals fails if the attacker has a power advantage of more than 15\unit{dB}, and the attacker is transmitting different navigation messages. 
However, this 15 dB limit is a limit defined by our signal processing.
Peripherals like multiple directional antennas and receivers can extend the 15\unit{dB} limit. 
Moreover, an attacker transmitting with more than 15\unit{dB} of power advantage can easily be detected and localized by the receiver.

An attacker can cause a denial of service attack by transmitting multiple signals that can overload the system.
Even though \recvname can handle multiple peaks through an iterative cancellation process, it is prone to resource exhaustion as each iterative cancellation increases process overhead.
The spoofing detection technique that we have adopted also has some limitations. For example, it can reliably detect spoofing only if auxiliary peaks are visible.
However, as mentioned earlier, SemperFi can work with multiple spoofing detection techniques that do not rely on auxiliary peaks.
\section{Related Work}
In recent years significant work has been done in developing robust GPS spoofing countermeasures. The work that comes closest to ours is the spoof-proof GPS receiver~\cite{Eichelberger2020spoofproof} and the in-line GPS spoofing mitigation technique~\cite{ledvina2001line}. 
In~\cite{Eichelberger2020spoofproof}, the receiver uses maximum likelihood estimates after dampening the attacker signal to estimate the correct location.
The in-line GPS spoofing mitigation technique~\cite{ledvina2001line} implements an extended RAIM method to filter outliers and correlation peak distortion techniques to detect spoofing signals.
Both these works are incapable of distinguishing adversarial peaks and fail against strong adversaries such as a seamless takeover attacker.
Signal cancellation has been explored in the context of GPS signals in~\cite{moser2019digital}. In this work, the goal is to attack the receiver by attenuating a specific satellite.
Specifically, successive interference cancellation has been studied to eliminate the near-far problem associated with pseudolites~\cite{madhani2003application}.
The authors treat overpowering pseudolites as interference because despite the signal being legitimate, it is so powerful that signals from GPS satellites are buried under the noise-floor, and the objective is to remove interference from this single interference.
Whereas in our work, we focus on removing the adversarial signal.
Other existing mitigation techniques can be categorized as i) Hardware-level mitigation techniques, ii) Signal processing level mitigation techniques, and iii) Cryptographic solutions. 
McMillin \etal~\cite{mcmilin2015gps} present a single-antenna design that can provide GPS jamming mitigation by null steering toward an optimal azimuthal direction. 
Such a solution requires additional hardware and might not be useful in a multi-spoofer setup, as described in ~\cite{tippenhauer2011requirements}. 
Borio \etal~\cite{borio2017fresh} provide an interference cancellation technique for recovering from GPS jammers. 
This work statistically models GPS jamming signals, which aides in jamming signal removal. 
McDowell \etal~in~\cite{mcdowell2007gps} provides a digital spatial nulling technique for spoofer mitigation. Several cryptographic solutions have been proposed for securing navigation messages. 
In \cite{kuhn2004asymmetric, cheng2009authenticity}, the authors propose an asymmetric and hidden marker approach for securing civilian GPS signals from signal-synthesis attacks. 
In~\cite{wesson2012practical}, the authors propose an authentication scheme by incorporating digital signatures. 
All these cryptographic solutions, although they prevent signal spoofing attack, requires key distribution and management. 
It is important to note that GPS is a public service used by millions of devices worldwide. 
Deployment of these solutions requires serious modifications to existing GPS infrastructure, which is impractical. 
Furthermore, cryptographic countermeasures do not prevent against record and replay attack~\cite{papadimitratos2008gnss}.
Several spoofing detection schemes require extra peripherals like multiple antennas~\cite{montgomery2011receiver,bhamidipati2019gps,meurer2016direction}, which detect discrepancies in the angle of arrival of GPS signals.
GPS signals and location estimates are correlated with data from extra IMU sensors~\cite{jafarnia2012detection, wendel2006integrated, titterton2004strapdown, farrell1999global} for detecting GPS spoofing attacks using vector-based tracking. 
Extensive work is present that focuses on the use of EKF to aid in recovering from GPS glitches~\cite{tanil2016kalman, hajiyev2013robust}. 
ArduPilot has one such implementation. 
Our experiments found that a spoofer can avoid detection by controlling the introduced error in the positions. 
In~\cite{zhang2018strategies, nashimoto2018sensor}, authors show how an attacker can create signals to defeat Kalman filter-based detection algorithms and inject false sensor data.
However, GPS/IMU sensor-fusion based navigation~\cite{narain2019security} has been recently shown to be vulnerable to attacks against on-road navigation systems.
Several works~\cite{tippenhauer2011requirements, jansen2016multi} propose using multiple receivers to detect spoofing signals by comparing reported positions of several GPS receivers with their deployed constellation.

Researchers have also proposed spoofing detection schemes that correlate civilian GPS signals with military signals~\cite{psiaki2011civilian}, cross-validation of PVT solutions across multiple navigation systems~\cite{nighswander2012crossvalid} \eg GPS, GLONASS, Galileo, etc. 
In~\cite{jansen2018crowd}, the authors leverage a crowdsourced network to detect GPS spoofing attacks.
In~\cite{borhani2020deep}, the authors discuss the use of deep learning schemes for spoofing detection and propose a detection approach based on machine learning.
Works like SPREE~\cite{ranganathan2016spree} and vestigial signal detection~\cite{wesson2011evaluation} provides a spoofing detection approach based on identifying auxiliary peaks.
All the above countermeasures only perform spoofing detection and are incapable of autonomous recovery during the spoofing attack.

\section{Conclusion}
In this paper, we presented \recvname, a single-antenna spoofer signal eliminating GPS receiver that is capable of providing uninterrupted legitimate locations even in the presence of a strong adversary. 
We designed, implemented SemperFi in GNSS-SDR capable of real-time operations and evaluated it using various GPS signal traces, real drones and popular embedded platforms. 
We showed that \recvname is capable of identifying adversarial peaks by executing flight patterns less than 50 m long and recover the true location in under 10 s.
Finally, we release the implementation of our receiver design to the community for usage and further research.

{\normalsize \bibliographystyle{plain}
\bibliography{references}}
\newpage
\section*{Appendix}
\begin{figure}[h]
    \includegraphics{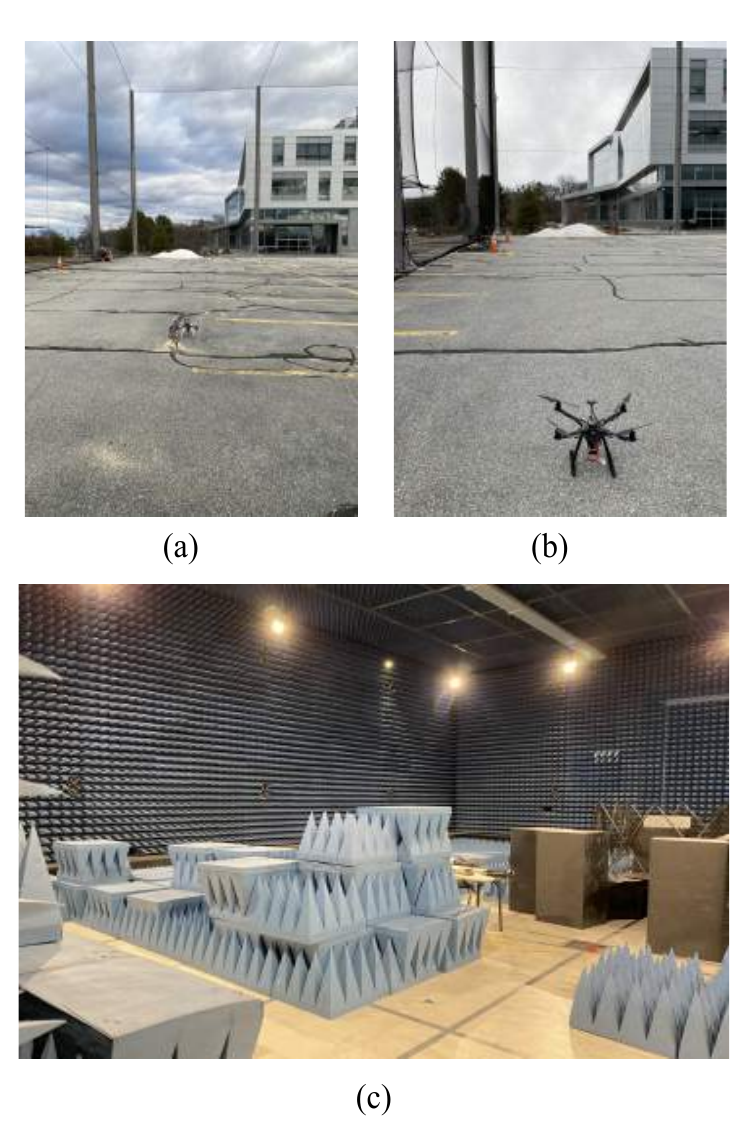}
    \\
    \noindent The drone testing and evaluation setup. (a) and (b) show the drones that we used (DJI Flamewheel F450 and Holybro S500) in the outdoor UAS testing facility. (c) shows the indoor anechoic chamber used for GPS spoofing and jamming experiments.  
    \label{fig:anechoic}
\end{figure} 

\end{document}